# Using Deep Learning to Detect Digitally Encoded DNA Trigger for Trojan Malware in Bio-Cyber Attacks

M.S. Islam[1*], S. Ivanov[1], H. Awan[4], J. Drohan[3], S. Balasubramaniam[2,] L. Coffey[3], S. Kidambi, W. Sri-saan[2]

**Abstract:** This article uses Deep Learning technologies to safeguard DNA sequencing against Bio-Cyber attacks. We consider a hybrid attack scenario where the payload is encoded into a DNA sequence to activate a Trojan malware implanted in a software tool used in the sequencing pipeline in order to allow the perpetrators to gain control over the resources used in that pipeline during sequence analysis. The scenario considered in the paper is based on perpetrators submitting synthetically engineered DNA samples that contain digitally encoded IP address and port number of the perpetrator's machine in the DNA. Genetic analysis of the sample's DNA will decode the address that is used by the software trojan malware to activate and trigger a remote connection. This approach can open up to multiple perpetrators to create connections to hijack the DNA sequencing pipeline. As a way of hiding the data, the perpetrators can avoid detection by encoding the address to maximise similarity with genuine DNAs, which we showed previously. However, in this paper we show how Deep Learning can be used to successfully detect and identify the trigger encoded data, in order to protect a DNA sequencing pipeline from trojan attacks. The result shows nearly up to 100% accuracy in detection in such a novel Trojan attack scenario even after applying fragmentation encryption and steganography on the encoded trigger data. In addition, feasibility of designing and synthesizing encoded DNA for such Trojan payloads is validated by a wet lab experiment.

# Introduction

Genetic sequencing has become an essential tool for analyzing numerous DNAs that are used in the field of medicine, agriculture, as well as forensics. Numerous systems have been developed over the years to increase accuracy, such as throughput shot-gun sequencing technologies (e.g., vector-borne pathogens detection in blood [30], food authentication and food fraud detection [31], or even molecular data to be transported through artificial biological networks [33] [34]). Recent developments in sequencing technology have also been miniaturized to allow mobile sequencing and one example is the *Minion* [29]. We have recently witnessed the importance of timely sequencing from oral samples due to the COVID-19 pandemic, which continues to apply pressure on the health care system [4]. The clear benefits of expanded COVID-19 testing [1] calls for an expansion of the existing testing (e.g. STEMI [2]) approaches. The importance of sequencing can also be seen in detecting and tracking mutations in other types of infectious diseases, where examples include Lassa Fever [3] or other prevalent pathogens [6], such as seasonal flu [5] or bacterial infections where new strains resistant to existing antibiotics can be identified [7][8].

[1]Walton Institute, Waterford Institute of Technology, Ireland. [2]School of Computing, University of Nebraska-Lincoln, Nebraska, USA. [3]Pharmaceutical & Molecular Biotechnology Research Centre, Waterford Institute of Technology, Ireland. [4]Munster Technological University, Ireland

[*] M.S. Islam is the corresponding author for the manuscript e-mail: sibleeislam@gmail.com.

As the genetic sequencing will inevitably introduce additional pressure on the already overburdened healthcare services, it is likely that the genetic analysis may be outsourced to private sequencing services. Similar approaches have already been successfully adopted for other testing programmes (e.g. Cervical Screening Programme in Ireland [9]). The services will act as an on-demand genetic-testing infrastructure that receives and analyses samples on behalf of the hospitals, medical practices and other healthcare organizations. While this approach alleviates pressure on the healthcare system, the system is vulnerable to Bio-Cyber Hacking [10].

Our definition of Bio-Cyber Hacking refers to an attack that is hybrid between ICT systems and biological mediums. From the ICT system side, we assume that the pipeline of the sequencing service uses a DNA-analysis toolbox infected with Trojan Software. Malware, such as a trojan, can be implanted at the API levels [26], within mobile software [27] and even in machine learning models [24]. Trojans can also be implanted into hardwares [20-22] of computers, as well as IoT devices [25]. In our scenario, the Trojan contains an empty slot for the IP address and port number for remote connections to an external machine. On the biological side, an attacker encodes the IP address and port number into DNA strands. Using DNA-steganography, the attacker devises synthetic DNA that contains the payload and still maintains resemblance with natural DNAs. We will explain the process in **Fig. 1,** where we will first explain a sequencing process for normal DNA (steps 1 - 3) and then explain a hacking situation (steps 4 - 8). In (**Fig.1 (1)-(2)**), the service uses one of the state-of-the-art sequencing techniques, e.g. shotgun sequencing, to analyze DNA materials extracted from each of the samples (e.g. E.Coli Plasmid and Cellular DNAs). The machine randomly splits DNA molecules into multiple fragments or reads of a predefined length, then it concurrently sequences each read to establish its nucleotide structure. The original DNA is then assembled from the reads (**Fig.1 (3)**). This is a computationally complex process that often involves the use of dedicated resources, often called DNA-sequencing pipeline [12]. Let us now consider an attack situation. Initially the Trojan remains dormant, while the toolbox performs the legitimate DNA-analysis. The trigger sample is collected by the hospital (i.e., by swabbing) and sends the samples to the sequencing service for analysis (**Fig.1 (4)**). The samples are then analyzed by the sequencer (**Fig.1 (5)**). There the sample is fragmented, sequenced and assembled (**Fig.1 (6)**). During the assembly, the DNA toolbox retracts the payload and wakes the Trojan (**Fig.1 (7)**), and this happens is when the DNA sample that contains the web address and port number of a remote server controlled by the attacker is detected by the digital DNA data that is passed from the sequencer to the computer that contains the DNA-analysis toolbox infected with the Trojan. The Trojan establishes a connection with the remote server (**Fig.1 (8)**), where the Trojan either opens a cyber backdoor, transfers files, or executes commands from the attacker. Either of these actions presents a substantial threat to the integrity of DNA-analysis and patient diagnostics.

In this article, we develop a solution that is complementary to the existing general-purpose techniques. The solution builds on our previous work that only focused on steganography techniques to hide IP address and port numbers into DNA strands [13] and investigates the use of input control (**Fig.1 (9)**) as a countermeasure to the Trojan Bio-Cyber attacks. The input control is an intermediary between the DNA-sequencer and the pipeline. With the help of a specially designed and trained Deep-learning Neural Network (DNN), the control assesses each DNA read generated by the sequencer to establish whether the read comes from a trigger sample. Absence of suspicious reads assures cybersafety of further DNA-assembly, but a detection of a trigger sample terminates its further processing. This prevents activation of the Trojan software and limits the pipeline's exposure. In recent times, there is a lot of interest in the use of deep learning for malware detection [19] [17] [18]. Deep learning techniques are also applied to Trojan detection [20, 23] in conventional cyber attacks. Finally, we validate whether the development of the DNA sequence of the payload of such a Trojan is realistic or not by conducting a wet lab experiment.

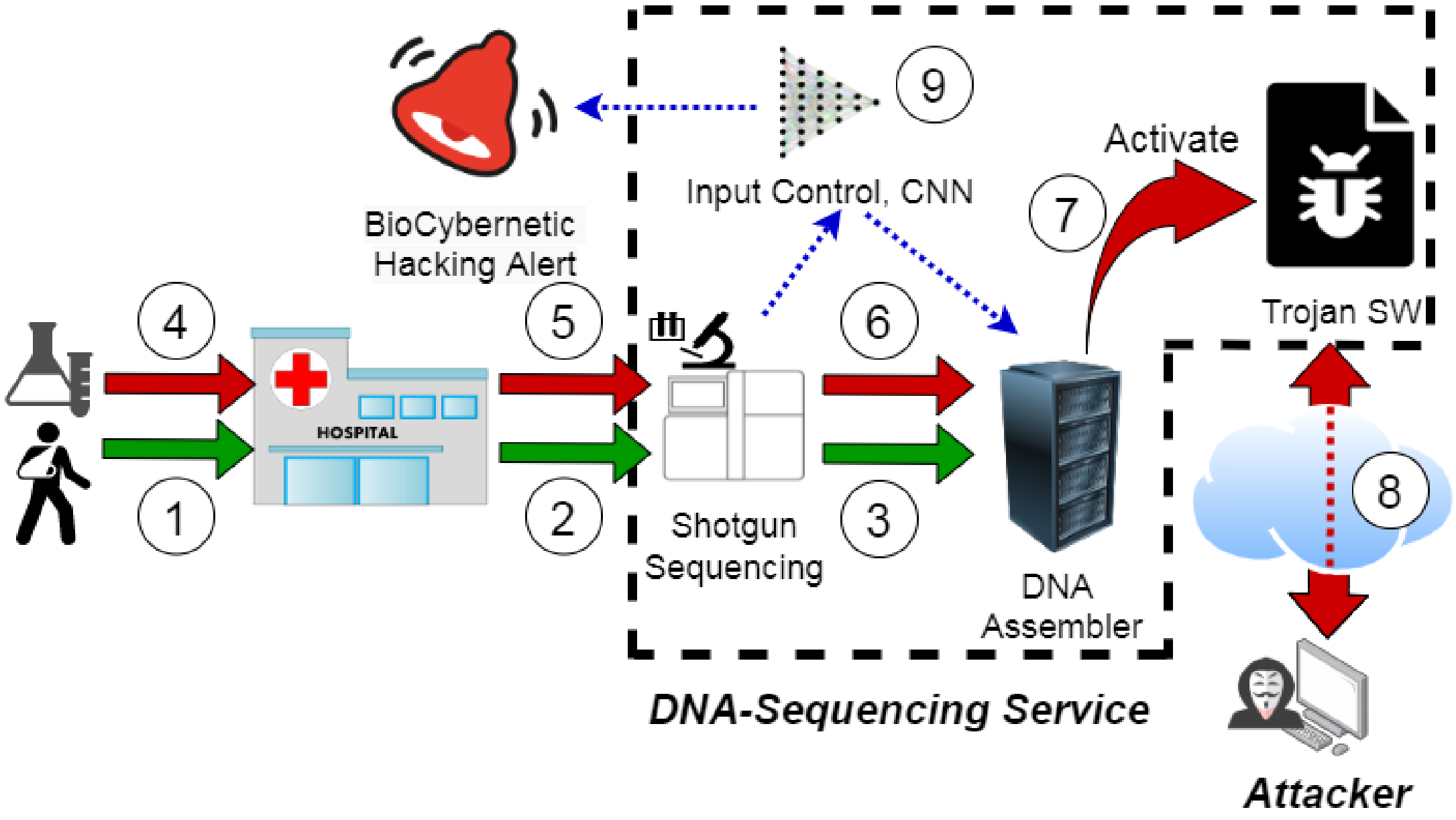

**Fig. 1:** *Hybrid Trojan Bio-Cyber Hacking Attack. Steps 1 - 3 indicate a typical genetic sequencing operation for patients. Steps 4 - 6 indicate a situation where a hacker has embedded their IP address and Port number into a DNA that will trigger a remote connection from a Trojan-horse infected software tool leading to a connection to the attacker in Step 8. Our proposed approach utilizes Deep-Learning to detect Trojan payload in digital data using encoded into DNA strands that can prevent the attack.*

**Fig. 2** illustrates the construction of the payload that is embedded into a DNA sequence, and in this specific example we focus on a bacterial plasmid. We re-designed the construction of the payloads to make it similar to a natural DNA sequence in order to increase detection difficulty. The construction of the DNA is based on the sequence used in [11].

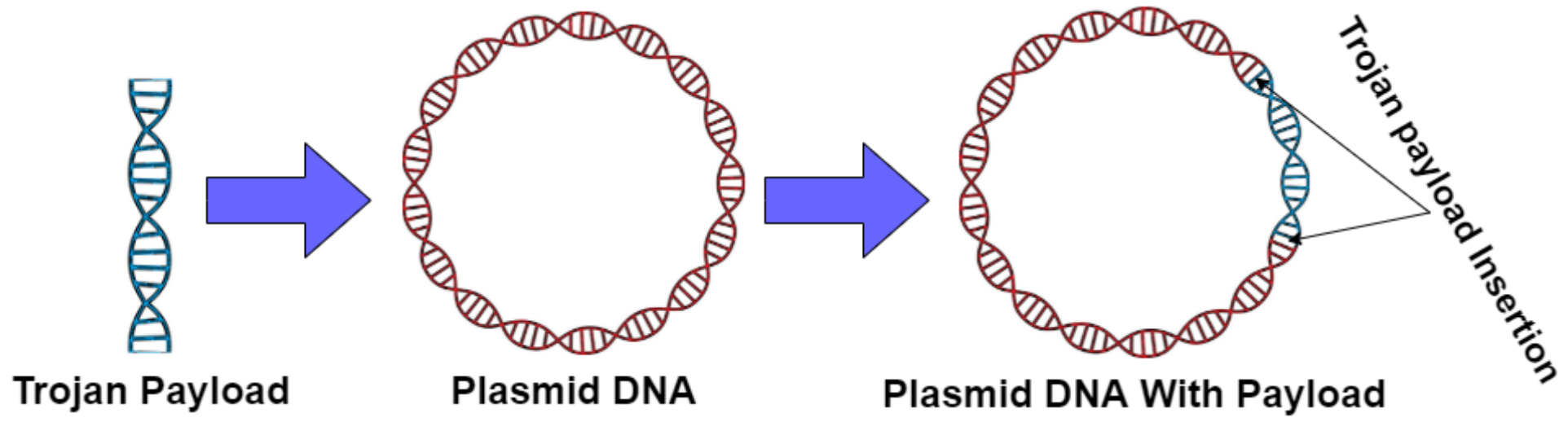

*(a)*

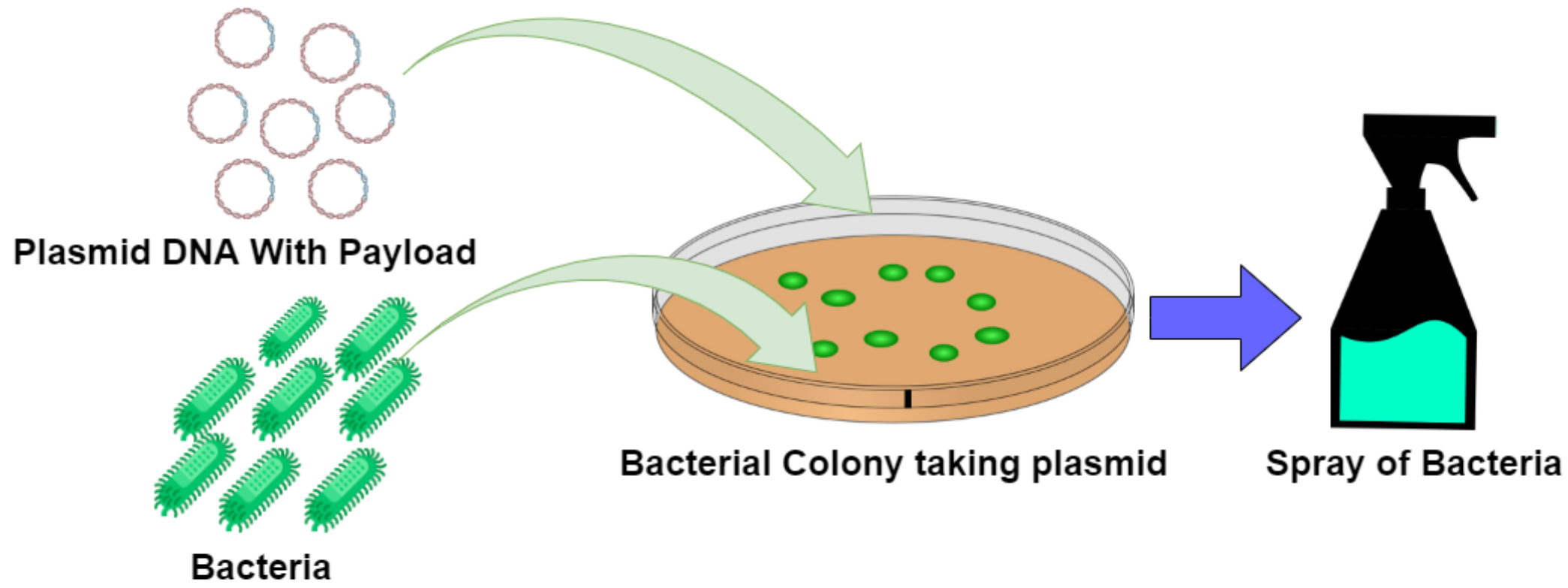

*(b)*

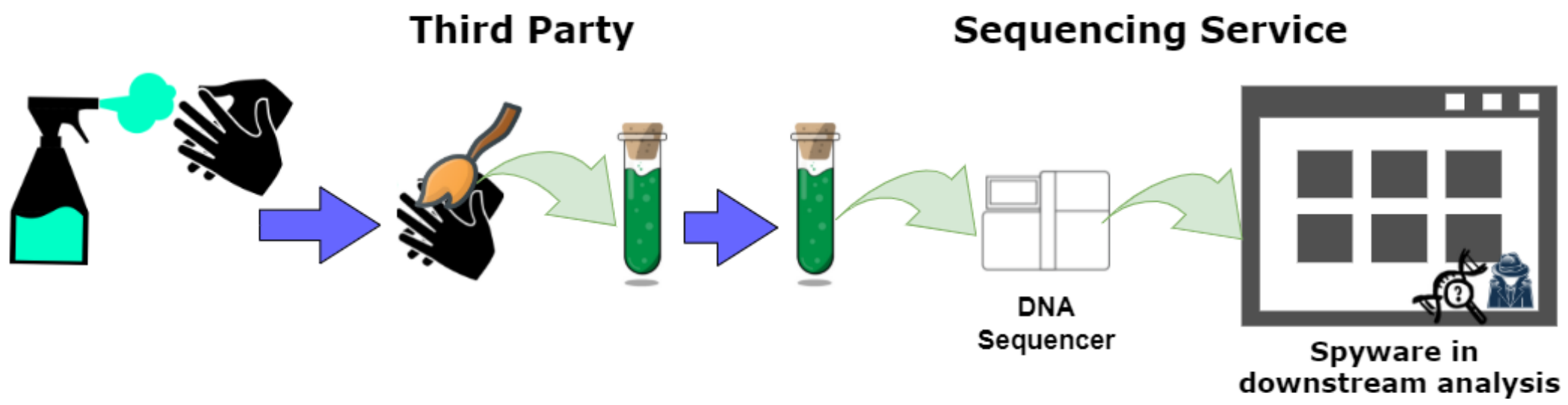

*(c)*

**Fig. 2: Trojan Bio-Cyber Hacking: Payload Preparation and Attack Scenario example using DNA plasmids.** ***(a)*** *A Trojan payload (using encryption and steganography) is encoded into a DNA sequence which is developed and inserted into the plasmid DNA. Antibiotic resistant gene sequences will also be inserted into the plasmid DNA in a similar way.* ***(b)*** *The DNA plasmid and the bacteria will be transferred into rich media so that the bacteria can uptake these plasmids [13]. Bacteria resistant to the antibiotic will survive and be transferred into a spray.* ***(c)*** *The bacteria can now be sprayed on hands or gloves and provided to a third party which can collect samples (from hand or gloves). The third party will then send these samples to the*

*company for sequencing. When the sequence will be processed by the tools having the Trojan, it will be activated to perform the malicious activities.*

# Methods

In this section, various terms used in the article will be defined and then the steganography techniques will be described, which is applied on the payload used for malicious activities as a means of secrecy of operations. Following that we will describe the injection method of the payload into a host DNA. This is followed up with the description of the deep learning model proposed as a detection method to counter the trojan attacks.

## Trojan payload

The payload DNA for triggering the Trojan malware will be encoded into a DNA sequence and will be referred simply as 'payload' in the rest of the article. The payload will be hidden inside a longer DNA string, which is considered as 'host DNA'. In order to prevent detection, the content of the payload will be first divided into smaller parts and then encoded into smaller DNA sequences, which will be called as 'fragments' and this process will be known as 'fragmentation'. The fragments can be inserted in a random order and at random positions of the host DNA. Substitution technique, i.e., replacing a nucleotide of the host DNA with a nucleotide of the payload DNA or fragment DNA (if fragmentation is applied), is considered as an insertion technique. 'Retention' is the process of skipping a particular number of nucleotide positions of the host DNA to substitute by the nucleotide of the encoded/fragment DNA while performing the insertion. Both encryption and retention will be considered when steganography is applied, where the encryption process will be performed before the retention. The details of the processes including encryption will be described in the subsequent sections of the article. After completing the insertion process, the obtained DNA string is considered as the 'resultant DNA'.

In general the host DNA string will be significantly larger compared to the encoded DNA for the payload. Therefore, the Trojan software needs to perform processes such as identifying those fragments, applying decryption and decoding techniques before merging and rearranging them in order to activate the malware process to trigger the hacking operation. As a result, the trojan software should apply these processes to integrate the substrings to create the full DNA string as an additional task beside performing its normal functional tasks. The caveat of such an approach is that the computational complexity will be significantly high and the trojan software might be under suspicion straight away as it will take significantly higher time and consume higher memory. To prevent this suspicious behaviour, the trojan software will need to efficiently determine the location to perform decryption and decoding and this will be achieved through 'tags'. The tags are tiny snippets of chosen DNA sequences that indicate the start and end of the fragments that will be searched by the trojan software, and we refer to this process as 'tagging'.

One of the critical challenges in packaging the Trojan payload is the delivery system which can act as the carrier for the DNA materials. To this extent, liposomes and lipid-based nanoparticles have been extensively used for targeted gene delivery to various coordinates. Liposomes, also referred to as vesicles, are extremely versatile carriers that have been studied and utilized extensively for drug delivery applications including gene and mRNA due to their ease of creation, large protective hydrophilic inner cavity for encapsulation, high degree of freedom for exterior customization, and controllable drug release kinetics. Recent success of mRNA vaccines for COVID  is attributed to such lipid based platforms as a delivery vehicle for mRNA. These can be extended to packaging the Trojan payload to enhance the stability of the DNA and also establish targeting capabilities to target specific locations for Cyber-hacking. Furthermore, there are innovative and robust platforms that can integrate these lipid nanoparticles embedded within substrate and matrix based on polymer based films that can control the release of these Trojan payloads and extend their stability [32]. Also this platform can also facilitate hiding these Trojan payloads from detection and embed multiple payloads. This platform provides ways to transport the Trojan Payload into the targeted region beyond security measures by embedding them into entities including clothes, skins, pens or papers as examples.

## Steganography

In this article we consider a scenario where the perpetrator encodes the attack details (i.e., web address and port number) into a DNA, which are used as a trigger sample. To avoid the detection of this sample and cover the identity of the attacker, the encoding uses an extension of the DNA Steganography technique proposed in [11].

The extended steganography technique proposed in this article has five steps and this includes *fragmentation*, *encryption*, *encoding*, *tagging* and *retention*. First, the web address and port number injected into the DNA are divided into fragments of a predefined length. Since each fragment is shorter than the original address, this will increase the difficulty in the detection process post injection. Next, the binary of the fragment is XOR-encrypted using a predefined key. This is followed up by encoding with four basic nucleotides, i.e., “00” bit-pairs are encoded as “A”, “01” as “C”, “10” as “G” and “11” as “T”. The ACTG-encoding (represent four nucleic acids, which are Adenine, Cytosine, Thymine and Guanine) is enclosed in the nucleotide brackets where the ACTG tags mark the beginning and the end of the injection within the DNA. These tags are selected so that the natural DNAs are unlikely to include both the start and end tags separated by a number of nucleotides that is required to encode a malicious fragment. The tags need to be sufficiently short in order to reduce the footprint of the injected fragment as well as increase the similarity with the host DNA and avoid detection. Finally, the retention stage expands the result of the tagging using the symbol “*” (see Eq. 1). The expansion is performed in a way that a predefined number of retention symbols is inserted between each of the two consecutive nucleotides. The positions of the retention symbol determine that the nucleotides of the host DNA will remain unchanged as a result of the malicious code injection. Thus, for a retention number equal to 2, only the first of each 3 consecutive nucleotides of the host DNA will be replaced. The second and third nucleotides will remain unchanged. This is done to increase the similarity between DNA of the trigger sample and the host DNA.

## Injection Methods

In this article we consider substitution as the preferred method of injecting the trojan payload into the host DNA. Consider the case when the trojan payload $d_{load}$ (with encoded nucleotides and retention symbol "*" after applying encryption and steganography as described above) is injected into the DNA, $d_{host}$, at position $i$. The result of the injection will present a nucleotide string $inj$, having the length equal to the length of $d_{load}$. The length of the $d_{host}$ and $d_{load}$ strings is determined by a function called $len$, which reflects the number of characters in both strings. The nucleotide at position $j \in [0,\ len(d_{host}))$ of $inj$ will be the insertion position $i$ and based on $d_{load}[j]$. If the value of $j$ does not fall between the range required for the injection position, which is from $i$ to $i + len(d_{load}) - 1$ as this location is required for the payload injection, then the actual nucleotide of host $d_{host}[j]$ will be used, i.e. $inj[j] = d_{host}[j]$. Otherwise, the value of $inj[j]$ will depend on $d_{load}[j - i]$, since the value $[j - i]$ determines the index of the $d_{load}$ and this has to be considered when it starts from 0 (for the very first substitution point $j = i$) up to $len(d_{load}) - 1$. If the $d_{load}[j - i]$ contains a retention symbol " * ", i.e. $(d_{load}[j - i] == " * ")$ then $inj[j] = d_{host}[j]$ (this means the original nucleotide is used for retention) otherwise $inj[j] = d_{load}[j - i]$ . This substitution procedure can be defined as:

$$inj[j] = \begin{cases} d_{host}[j] \text{ if } j < i \text{ or } j \geq i + len(d_{load}), \\ d_{host}[j] \text{ if } j \in [i,\ i + len(d_{load})) \text{ and } d_{load}[j - i] == " * ", \\ d_{load}[j - i] \text{ if } j \in [i,\ i + len(d_{load})) \text{ and } d_{load}[j - i] \in \{A, C, T, G\}. \end{cases} \quad (1)$$

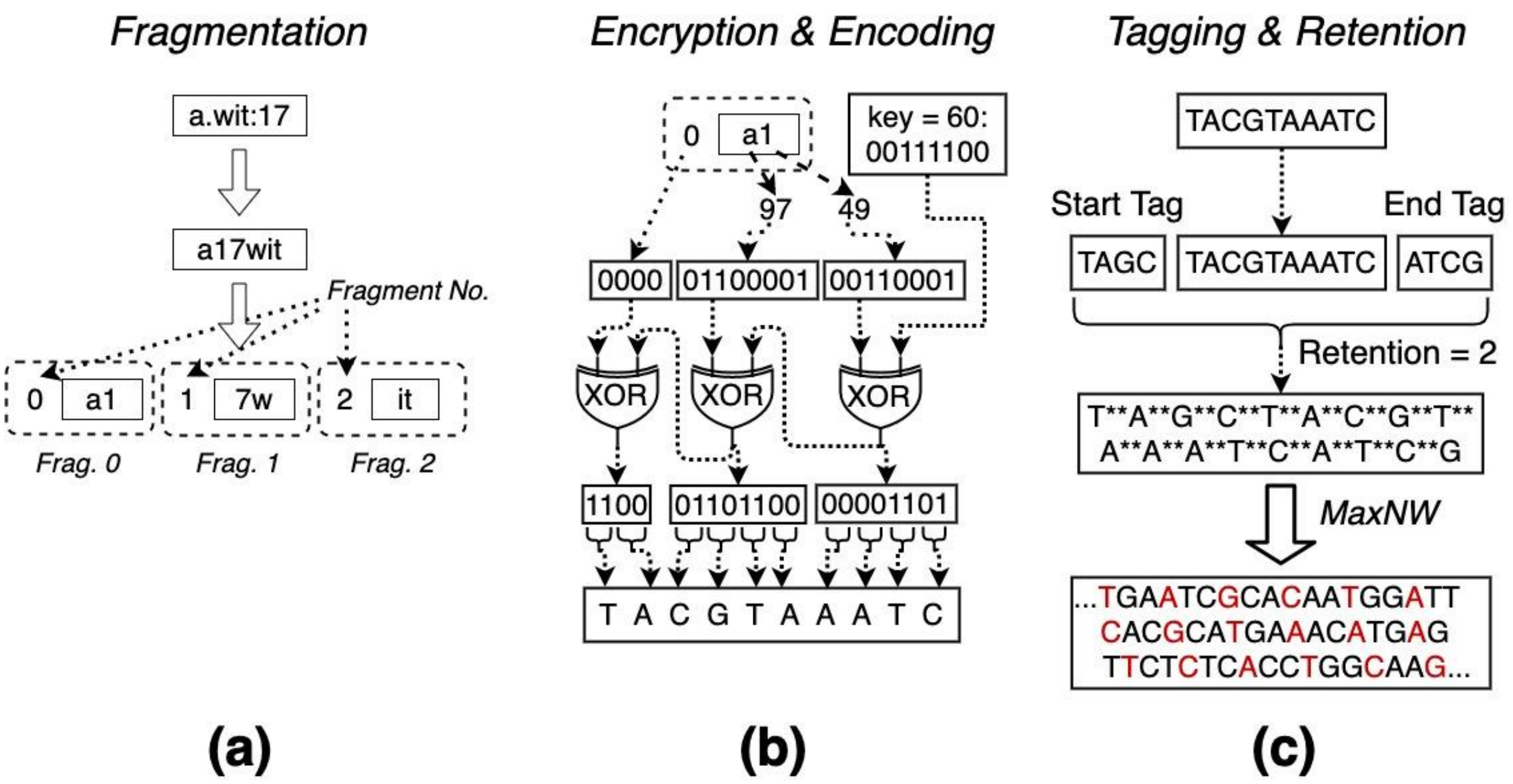

**Fig. 3:** *DNA Steganography, Workflow: (a) payload fragmentation, (b) fragment encryption and encoding, (c) tagging, retention and host injection.*

We define *elementary* domain $dom_{ELM}$ that consists of all the possible positions for a trojan payload injection. Naturally, such a substitution can be carried out only from the position $i$ onwards and is represented as:

$$dom_{ELM} = [0,\ len(d_{host}) - len(d_{load}) + 1], \qquad (2)$$

which is referred to as the *injection domain* and refers to the indices (i.e., values of $i$) of $d_{host}$. We use $inj(d_{host}, d_{load}, i)$ to denote the substitutions in $dom_{ELM}$. Similarly, to denote the substitutions carried out on subdomain $dom \subset dom_{ELM}$, we use $inj_{dom}(d_{host}, d_{load}, i)$. This subdomain introduces additional restrictions that may be required to preserve particular areas within the host DNA. **Fig. 3** presents the five stages/steps involved in the DNA steganography technique used in this article.

Note that in this article we only consider payloads that consist of a web address (represented by a Tiny URL) and port number of a remote server controlled by the attacker[2]. The payload has the following semantics:

<**prefix**: character string>**.**<**suffix**: character string>**:**<**port number:** string of digits>

As mentioned above, the fragmentation (**Fig. 3(a)**) is the first stage of the DNA steganography. First, the payload is rearranged so that the address prefix is followed by the port number and then the address suffix. This representation allows the reduction in the auxiliary "." and ":" characters from the payload, and therefore, size reduction of the entire payload. Subsequently, the rearranged payload is divided into fragments, substrings of a predefined length (e.g. 2 characters as shown in Figure 3). Each of the fragments is attached with its serial number as a prefix. As only tiny URLs are used in the tojoan payload address, we assume that no more than 16 fragments can be formed.

[2] For payloads used in the analysis, please see: https://bitbucket.org/sibleeislam/bio-cyber-hacking

The next step after frangementation is encryption, where each fragment is encrypted and nucleotide-encoded as illustrated in **Fig. 3(b)**. At this stage, the fragment is represented as a bit-array where the first 4 bits represent the fragment's serial number, followed by a series of 8-bit representations of fragment characters. Each character is represented by the binary of its ASCII code. The bit-array is then XOR encrypted using a predefined key (e.g. 60 as depicted in **Fig. 3(b)**). This results in a sequence of bit-pairs, which are then encoded into nucleotides strings that represent the DNA.

The next step after encryption is encoding as shown in **Fig. 3(c)**. The nucleotide-encoding of the fragment is attached with a start and end tag as prefix and suffix, respectively. The resultant string is then expanded so that a predefined amount of retention symbols is added between each two consecutive nucleotides (e.g., 2 symbols as in **Fig. 3(c)**). The expanded string is then injected into the host DNA using MaxNW procedure, which is described next.

## MaxNW Technique

Needleman-Wunsch, or NW score is one of the most popular methods to assess the similarity between two DNA samples. This score considers the string-based nucleotide representation of the DNA molecules and calculates the number of symbol substitutions, gaps (i.e., symbol insertion or deletion) and their expansions (i.e., continuation of gaps) required to align two strings. Depending on the circumstances, a specific penalty system is applied to each of the operations as well as matches between DNA nucleotides. The system is constructed in a way to favor certain alignment patterns. As in the experiments performed in this work, injecting payload typically constitutes not more than 10% of the host DNA string size, therefore we use PAM10 substitution scoring matrix [28] as the cost matrix for nucleotide substitution. Following this methodology outlined in [15], we set the costs for the gap opening and extension to 15.79 and 1.29 for the PAM10 substitution, respectively.

In this article, we use NW scores to measure the similarity between $d_{host}$ and $inj(d_{host}, d_{load}, i)$. Based on the penalties defined above, the NW score increases as similarity between $d_{host}$ and $inj(d_{host}, d_{load}, i)$ increases and reaches its maximum if $d_{host}$ and $inj(d_{host}, d_{load}, i)$ are equal. In other words, the injected payload fits into the $d_{host}$ naturally at position $i$. Lets assume the NW score is maximum when the insertion position (the value of $i$) is $i_{max}$. To emulate the attacker, the malware NW score, $MaxNW_{dom}$, is defined as:

$$MaxNW_{dom}(d_{host},\ d_{load}) = inj(d_{host}, d_{load},\ i_{max}), \tag{3}$$

where

$$i_{max} = Arg\Big(max_{i \in dom} NW\big(d_{host}, inj(d_{host}, d_{load}, i)\big)\Big). \tag{4}$$

When multiple payloads for malicious activity injections $D_{load} = \{d_{load,1}, \dots , d_{load,n}\}$ are introduced into the same host DNA, dynamic programing is used to determine the optimal positions for the injections. The technique employs a recursive procedure, where at each step the best insertion is sought amongst all possible positions. So, initially $inj(d_{host}, d_{load}, i)$ and $dom_{ELM}$ are considered for the substitution and the domain for the substitutions for each of the payloads. Then the injection position of the payload having maximum NW Score will be considered for that particular payload injection and that portion of the injection will be restricted for further injections. For further steps, the subdomain $dom$ and injection for subdomain $inj_{dom}(d_{host}, d_{load}, i)$ will be considered as the restriction is applied. Lets assume, the maximum NW Score and the indices considering subdomain are $MaxNW_{dom^*}$ and $, i\ ^*$ respectively. The injection process will be repeated until all the payloads are injected. Thus, this recursive procedure can be described as:

$$MaxNW_{dom}(d_{host}, D_{load}) = MaxNW_{dom^*}\Big(MaxNW_{dom}\big(d_{host}, d_{load,i^*}\big), D_{load}/d_{load,i^*}\Big), \tag{5}$$

where

$$i^* = Arg\Bigg(max_{j \in [0, len(D_{load}))}\ NW_{dom}\big(d_{host}, d_{load,j}\big)\Bigg). \tag{6}$$

## Deep Learning

In this article, we use a 1-Dimensional Convolutional Neural Networks (1D CNNs) to identify the trojan payload within the natural DNAs. This section will provide a brief overview of the CNNs we utilized for this work. An overview of various Deep Learning methods, including CNNs, used in genetics analysis can be found in [16].

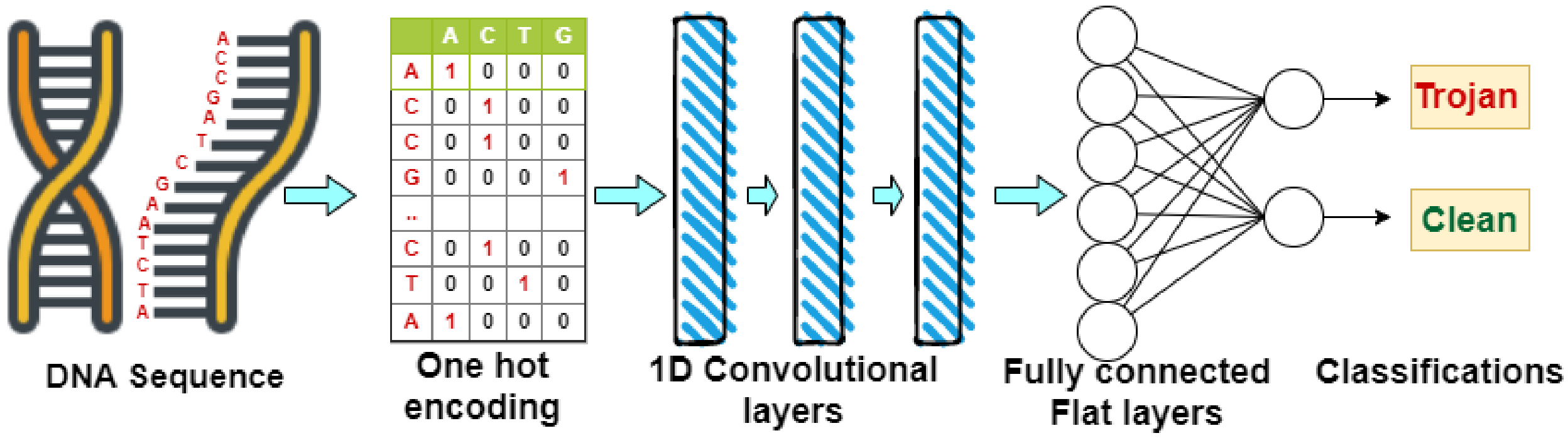

**Fig. 4:** *1-Dimensional Convolutional Neural Network (1D CNN): Architecture*

**Fig. 4** depicts the typical architecture of a 1D CNN. Similar to any other neural network, the 1D CNN consists of neurons organized in layers. The architecture proposed in this article uses the following layers: input, convolution, pooling, and dense.

The first layer represents the input of the network. Here, each of the DNA sequences' classification is transformed into the set of primary features, i.e., inputs of the network. Each nucleotide of the DNA is represented by a vector of 5 boolean indicator values. The first 4 values indicate whether the nucleotide are found to be equal, whereas the 5th value indicates whether the nucleotide can be determined (i.e. N - undetermined). As an example, A-nucleotides of the DNA will be represented by (1,0,0,0,0) indicator vectors, C-nucleotides will be represented by (0,1,0,0,0), and undetermined nucleotides will be represented by (0,0,0,0,1). To formulate the primary features of the entire DNA, indicator vectors for all its nucleotides are concatenated in the order of the pattern that is found in the original DNA.

The input layer is followed by a number of CONV1D layers as shown in Fig. 4. At each layer, multiple filters are applied to the kernels of a particular size. The resultant product is then subjected to ReLU activation. CONV1D layers are followed by 1 MaxPool, one dense layer with ReLU activation function, and finally 2-neuron SoftMAX layer, the output of which provides the certainty of the sample to be determined if it contains the address information. In this article, we consider networks with varying numbers of CONV1D layers, the size of their kernel and the number of filters used. We also investigated the impact of the kernel size of the MaxPool layer and the size of the ReLU dense layer. Each network is trained for 3000 epochs using 75% of all available DNA samples. The remaining 25% of the samples are used to test the performance of the trained network.

# Results and Discussion

For the Trojan infected softwares, the secrecy of operation is of paramount importance. The longer the Trojan remains undetected, the more extensive the damage it can cause. For the Bio-Cyber hacking attack considered in this article, it is of vital importance for the attacker to maintain a natural appearance of the trigger sample containing the address details. If we use an unnatural DNA structure as a part of the hybrid attack it can be flagged as suspicious not only by the detection method proposed in this article, but also by the similar less sophisticated versions of this system proposed in previous works [13].

In this section we begin the discussion by evaluating the possible actions of an attacker to design a natural trigger sample. We follow this up by investigating the accuracy with which these trigger samples can be detected by a CNN. Finally, we describe the wet lab experiments that were used to produce the DNA with the address, in order to validate the potential of creating such a DNA sequence that is used as the trigger sample for our attack.

## Trigger Sample Design

For this article we propose the use of *E.Coli* plasmids that will encode the address of the attacker. E.Coli bacteria have been sufficiently studied in literature and their plasmids can be synthesized and modified with relative ease. Once the attacker identifies a suitable DNA structure, E.Coli plasmids can be readily synthesized in various laboratories across the globe such as *EuroFins Genomics and Twist BioScience* [13]. In this section, we present the design of the plasmid DNAs that contains the trojan payload that will maintain the original E.Coli plasmids sequence. Specifically, we evaluate the use of DNA steganography (as described in the Methodology section) for injecting the address payload into an E.Coli plasmid (host) DNA to maximize similarity between the resultant $inj$ and host DNAs $d_{host}$.

This evaluation requires 1000bps reads randomly sampled from the plasmid DNAs made available via *AddGene* repository. The sampling serves two purposes. First, it mimics the operation of a DNA-sequencer (e.g., Roche 454 FLX+ [14]) that may be specifically targeted by the attacker. In this case, a higher number of DNA-reads produced by the sequencer (i.e., 700-1000 bps) will provide better cover for the trojan address payload and, thus, increase the chances for the hybrid attack to be successful. Secondly, the sampling can significantly increase the amount of DNA-data used in the evaluation, where we draw 4356 reads from 716 E.Coli plasmid DNAs stored in the AddGene repository.

Since the steganography technique has five key steps, the *encoding* step is fixed and cannot be varied, but the attacker is free to finetune the tagging, fragmentation, encoding, retention, and encryption steps. In **Fig. 5** we show the impact of different parameter combinations, e.g. size of the fragment, number of retention positions, and value of the encryption keys.

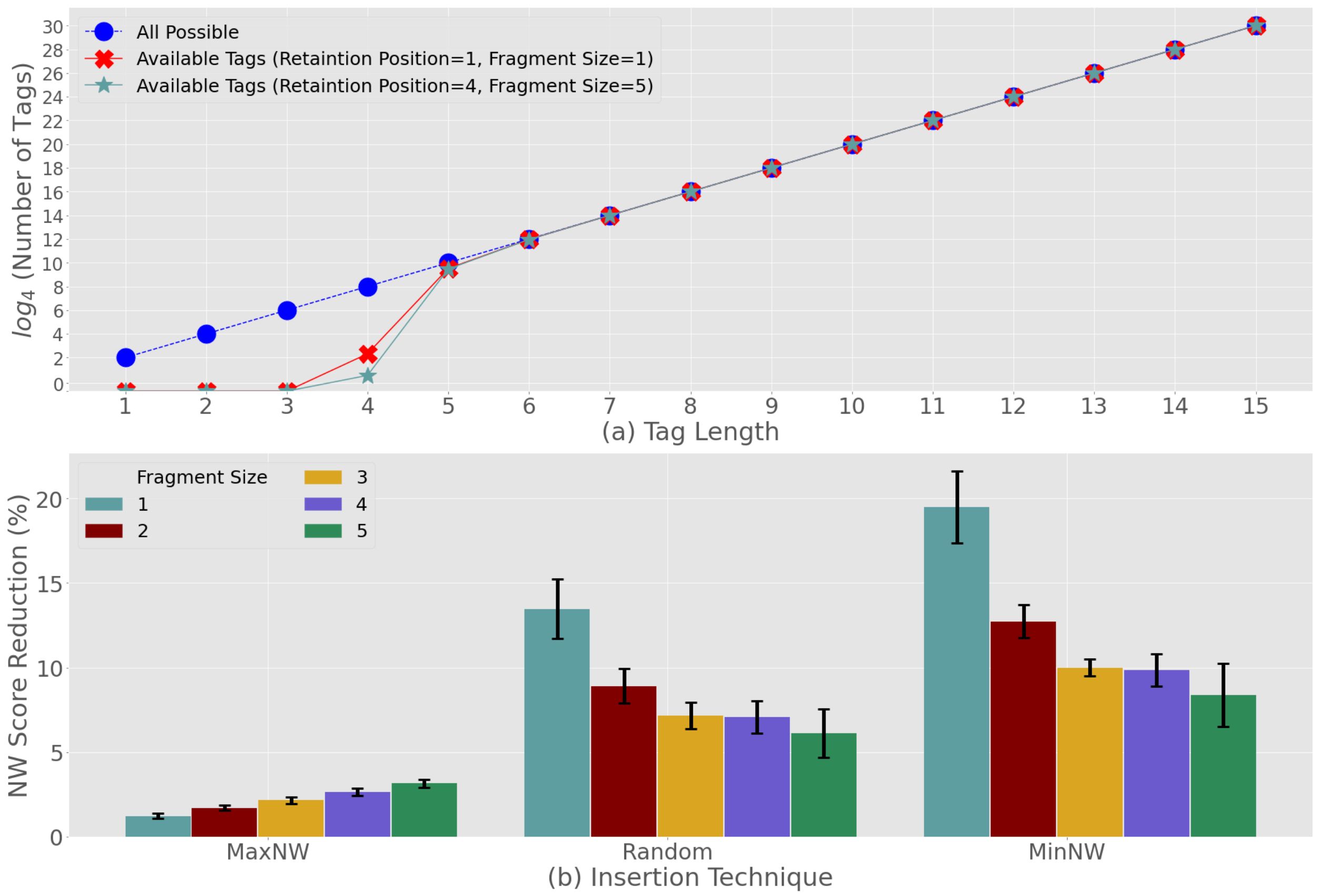

**Fig. 5:** *Trigger Sample Design with the use of DNA-Steganography: (a) nucleotide tag selection; (b) the impact of fragmentation.*

**Fig. 5 (a)** depicts the relationship between the length of nucleotide tags and their availability. The tags mark the start and the end of the trojan payload injections into a plasmid DNA. These tags that mark the start and the end of the trojan payload are two potentially different nucleotide sequences of the same length. The sequences are selected in a manner that a host DNA is unlikely to include both tags separated by nucleotides. Note that the number of these nucleotides are obtained directly from the fragment size and the retention (i.e. retention of host nucleotides) parameters of the steganography technique. The results in **Fig. 5 (a)** correspond to various values of these two parameters. From these results we learn that a predictable growth of tag availability is associated with the increase in tag length. As the number of all possible nucleotide sequences grows exponentially, it can overcome the number of unique sequences in genuine DNA reads for 4-nucleotide tags. We also realize that any further increase in the tag length (i.e., 5 and beyond) will make the number of unique sequences negligible, leaving the attacker with ample choice of nucleotide tags. The strength of this effect is such that it can be seen for all fragment sizes and retention values. As a result of this observation, we use a minimum 5-nucleotide tags for the remainder of this article as this is the lowest length that allows for the substantive tag availability.

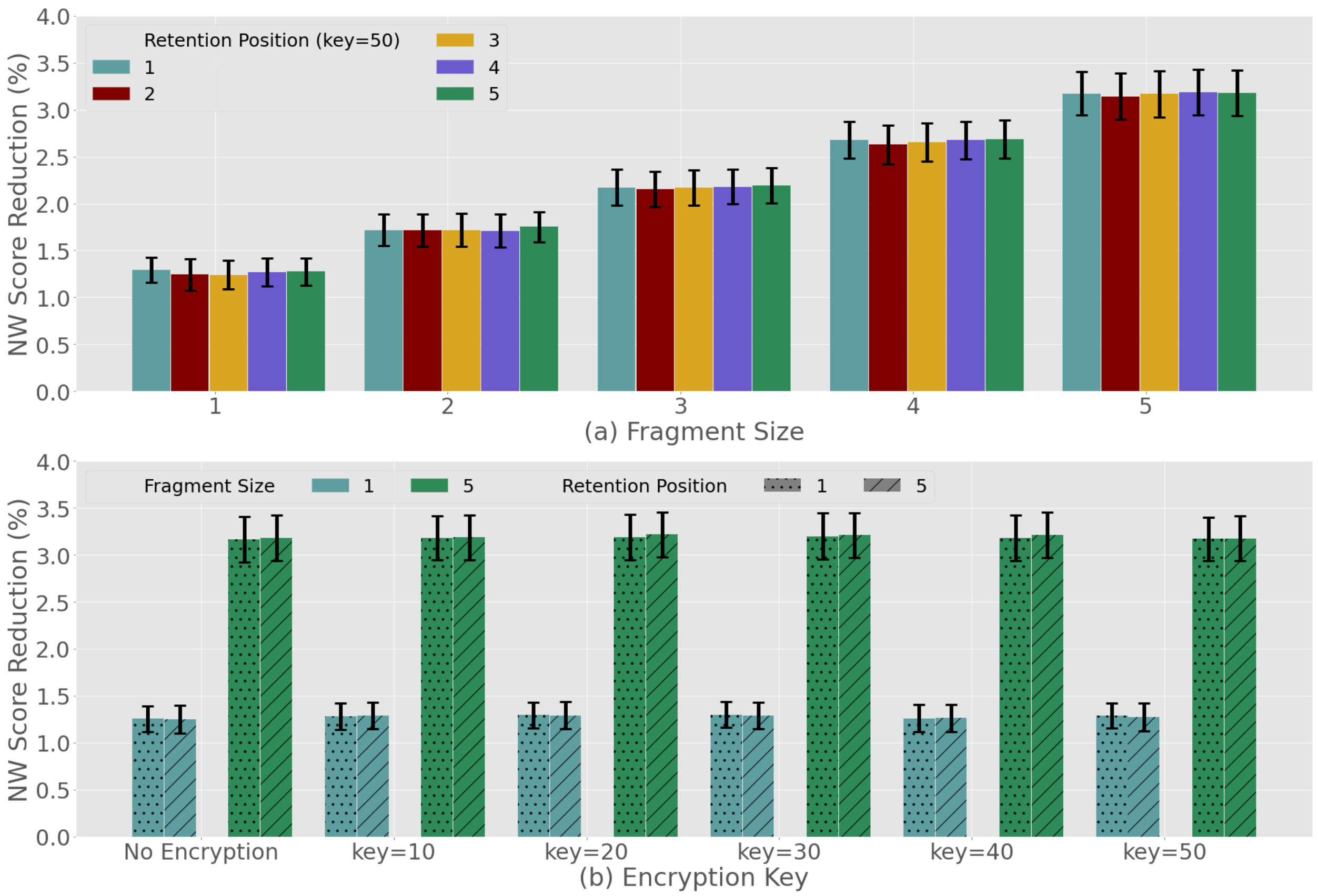

**Fig. 6:** *Trigger Sample Design, the use of DNA-Steganography: (a) retention of host nucleotides; (b) payload encryption.*

In **Fig. 5(b)** we study the impact of the fragment size selection on the similarity between the host DNA before and after the injection of trojan payload. This similarity is assessed by using Needleman-Wunsch (NW) scores (described in Methodology). The system is designed in such a way that the Needleman-Wunsch score grows as the similarity between the two DNAs increases. The value of this score is absolute maximum (i.e. MaxNW) when either the DNAs are identical, or the trojan payload address is inserted into the host DNA naturally. Since due to tagging this is not possible we use the maximum (i.e. the NW score between host the DNA and itself) value to benchmark the score reduction due to the payload injection. Furthermore, in order to ensure the optimal payload injection, the steganography uses MaxNW technique (described in Methodology). To demonstrate the efficiency of this technique, **Fig. 5 (b)** presents a comparison of performance with two alternative techniques, i.e., Random and MinNW. Random technique injects the payload at an arbitrary position through uniform distribution, whereas MinNW is a dynamic programming technique that seeks the worst possible injection position for a payload. This means that MinNW is a mirror-image of MaxNW which can minimize the score between the host and injected DNAs. This phenomenon is reflected in **Fig. 5 (b),** where MaxNW results in significantly lower score reduction compared to MinNW, whereas the score reductions by Random technique lies approximately in the middle of those produced by MaxNW and MinNW. From this we conclude that the MaxNW and MinNW techniques can show the whole range of score reductions that may occur due to payload injections. This also reaffirms that MaxNW is the best technique amongst all three possible techniques. In addition, a closer inspection of the results for the MaxNW technique also clarifies the impact of payload fragmentation. We realize that using a larger fragment size in the host DNA can effectively reduce the similarity between the host and injected DNAs.

Next in **Figs. 5(a)** and **(b)** we investigate the impact of different *retention* as well as *encryption* choices of the attacker. The results are presented only for MaxNW which is the optimal injection technique we have selected. For both the retention of host nucleotides or payload encryption, we realize that there is no significant effect on the NW score. In particular **Fig. 6(a)** shows no change in the NW score reduction can be attributed to different retention numbers for various fragment sizes for payload encrypted with a key equal to 50. **Fig. 6(b)** shows similar results, where payload fragments of 1 and 5 characters are injected using 1 and 5 retention numbers. For this case, we also observe  no change in the NW scores when encription keys are utilized. Based on these results, we can conclude that neither *retention* nor *encryption* are likely to disguise the trigger sample. Although we note that neither of these two steps can help the payload appear more naturally, however they still remain an essential part of the steganography process. This is because these steps play a key role in maintaining the anonymity of the attacker as they are designed to protect the payload (i.e. network address and port number), which may identify the attacker. For the case when a trigger sample is identified, the retraction of the payload will require knowledge of both the *retention number* and the *encryption key* used by the attacker.

## DNN Detection Accuracy

Although the natural appearance of the trigger sample is necessary to disguise the hybrid attack and avoid detection by less sophisticated methods (e.g. NW comparison with known DNAs), the trojan payload address injection may still be discoverable with the help of other techniques. In this section, we will explore this by evaluating the detection of trigger samples using a state-of-the-art Deep Learning approach. We achieve this by investigating the performance of a 1-Dimensional Convolutional Neural Networks (CNN). The results in **Fig. 7(a) and (b)** summarize the performance of various CNNs topologies with respect to the four hyper-parameters considered in this article. This includes, (i) the number of hidden *layers (1 and 2)*, (ii) the sizes of the *filter* (4, 8 and 16), (iii) size of the *kernel* (3, 5 and 8), and (iv) size of the *maxpool* (2 and 4) used in the network. The results are then obtained for trigger samples obtained from natural DNA using 0-retention and no payload encryption. This means that we can establish a baseline predictive capacity of CNNs and determine the most suitable network topology. This suitable topology is then further tested to evaluate the ability to cope with additional uncertainties introduced by nucleotide retention and payload encryption.

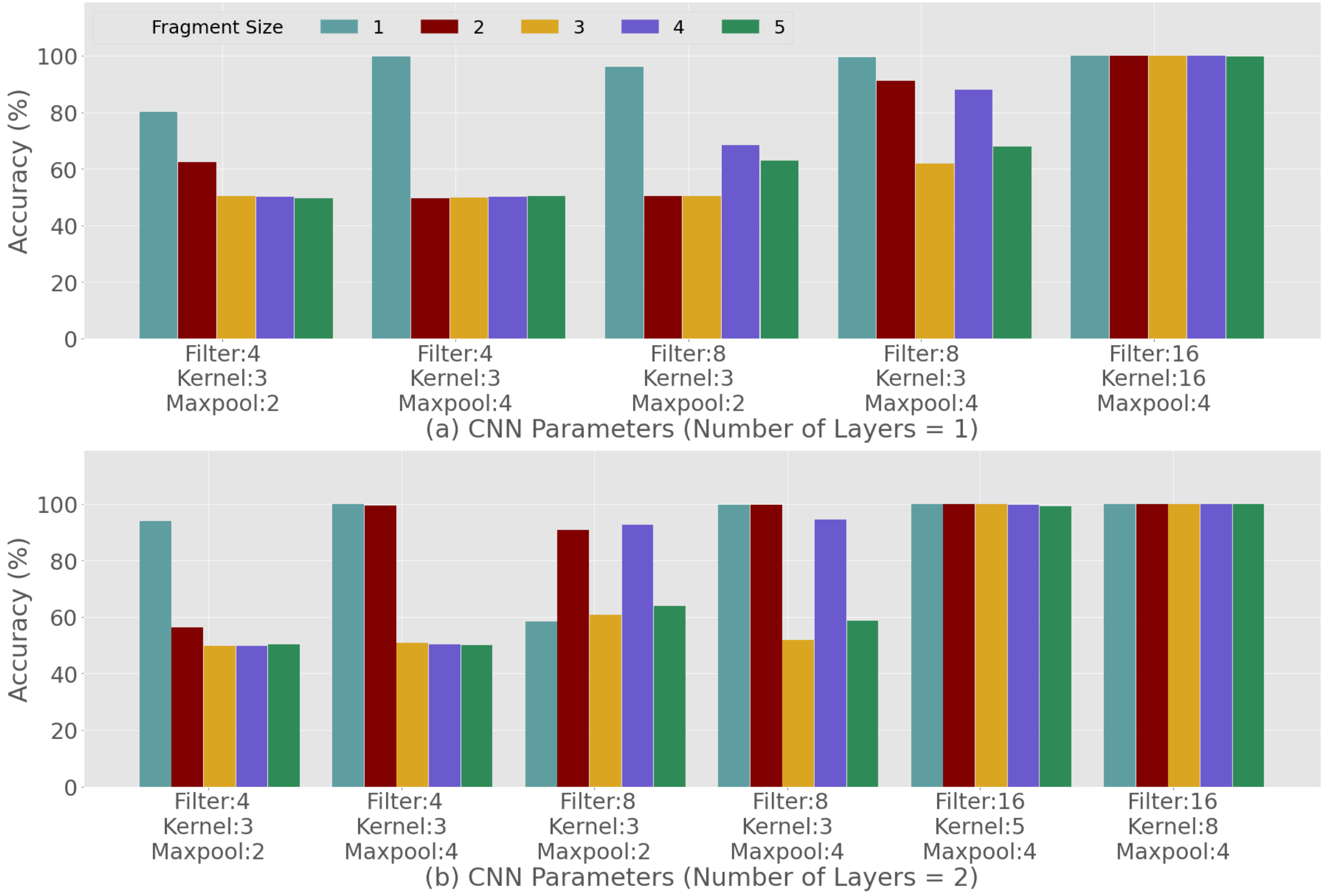

**Fig. 7:** *DNN-based detection of trigger samples amongst genuine E.Coli plasmids: hyper-parameter optimization (no encryption or retention) using (a) 1 and (b) 2 hidden layers.*

For this purpose, we simulated 180 scenarios for 36 combinations of hyper parameters and for 5 different fragment sizes, with no retention and no encryption. We obtain the best accuracy (99.9%-100%) for all 5 fragment sizes when we have 1 hidden layer, kernel size 16, 16 filters and 4×4 max pool size (**Fig. 7(a)**). Similarly, we obtain the best accuracy for the case we have an additional layer (2 hidden layers), 16 filters, kernel size 5 and 4×4 max pool (**Fig. 7(b)**). These features are mainly learned by the kernel, so larger kernels and higher number of filters result in achieving the best accuracy. However, in this article we prefer to use a smaller number of required hidden layers to increase the execution time performance. Therefore, for the rest of the experiment we consider the CNN topology with 1 hidden layer, kernel size 16,16 filters and 4×4 max pool.

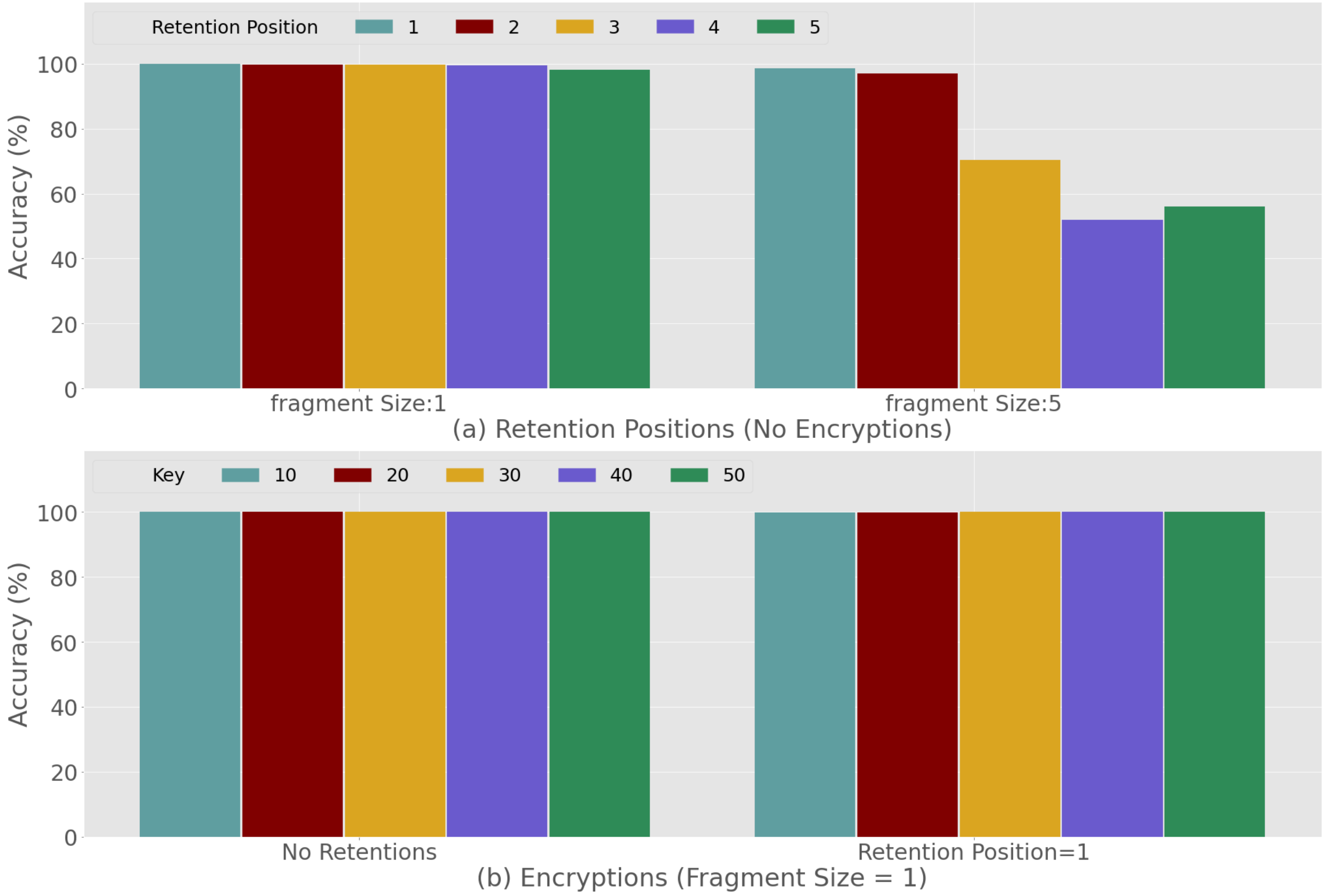

**Fig. 8:** *DNN-based detection of trigger samples amongst genuine E.Coli plasmids: the impact of nucleotide retention (a) without encryption and (b) with encryption and with prior knowledge of the encryption key***.**

Next in **Fig. 8** we analyze the impact of the fragment size, retention values and encryption on the trojan address detection. In particular, **Fig. 8 (a)** presents the detection accuracy for the highest and lowest fragment size values (1 and 5), and all the retention numbers (1 to 5), when no encryptions are applied. We made an assumption that if we split the payload into an increasing number of fragments it will be relatively easy to escape the detection. In such a case it will be comparably difficult to locate the complete trojan payload address and, therefore, be relatively harder to make sense out of a more tinier part of the payload. Furthermore, as shown and explained in the previous section (**Fig. 6(a)** and **(b)**), the DNA sequences remain much more natural for smaller fragment sizes. Based on this knowledge, a potential hacker might prefer to choose a smaller fragment size. However in reality this approach will leave more tags as low fragment size translates to increase in number of tags. Therefore, this approach can support the CNN model, which can learn from the tag patterns and the result in **Fig. 8 (a)** illustrates this.

On the other hand, in a real world scenario it will be a significant challenge to design an optimal model which can account for many variations of tags. Interestingly, we observe that for higher fragment sizes, the accuracies deteriorate very slightly until there is a higher retention number as well **(Fig. 8 (a))**. This indicates that the model proposed in the article does not completely rely on learning the tag patterns. Furthermore, the higher retention number means more number of nucleotides (from the original sequence inside the tags) which will result in more variations and harder detection. However, we note that for fragment size 1 the accuracies are very high for all retention numbers. Overall, the accuracies start to deteriorate significantly for the higher fragment sizes with higher retention numbers **(Fig. 8 (a))**. To analyze the impact of encryption on the trojan address payload detection, we consider fragment size 1 with no retention and retention size 1 as we obtain the best accuracy for these options. We apply encryptions with various key values ( *key*ϵ*{10,20,30,40,50})*. In **Fig. 8 (b),** the results show that there is no significant change in accuracy when applying various encryption keys. Please note that both the training and test data are using the same key value for encryption.

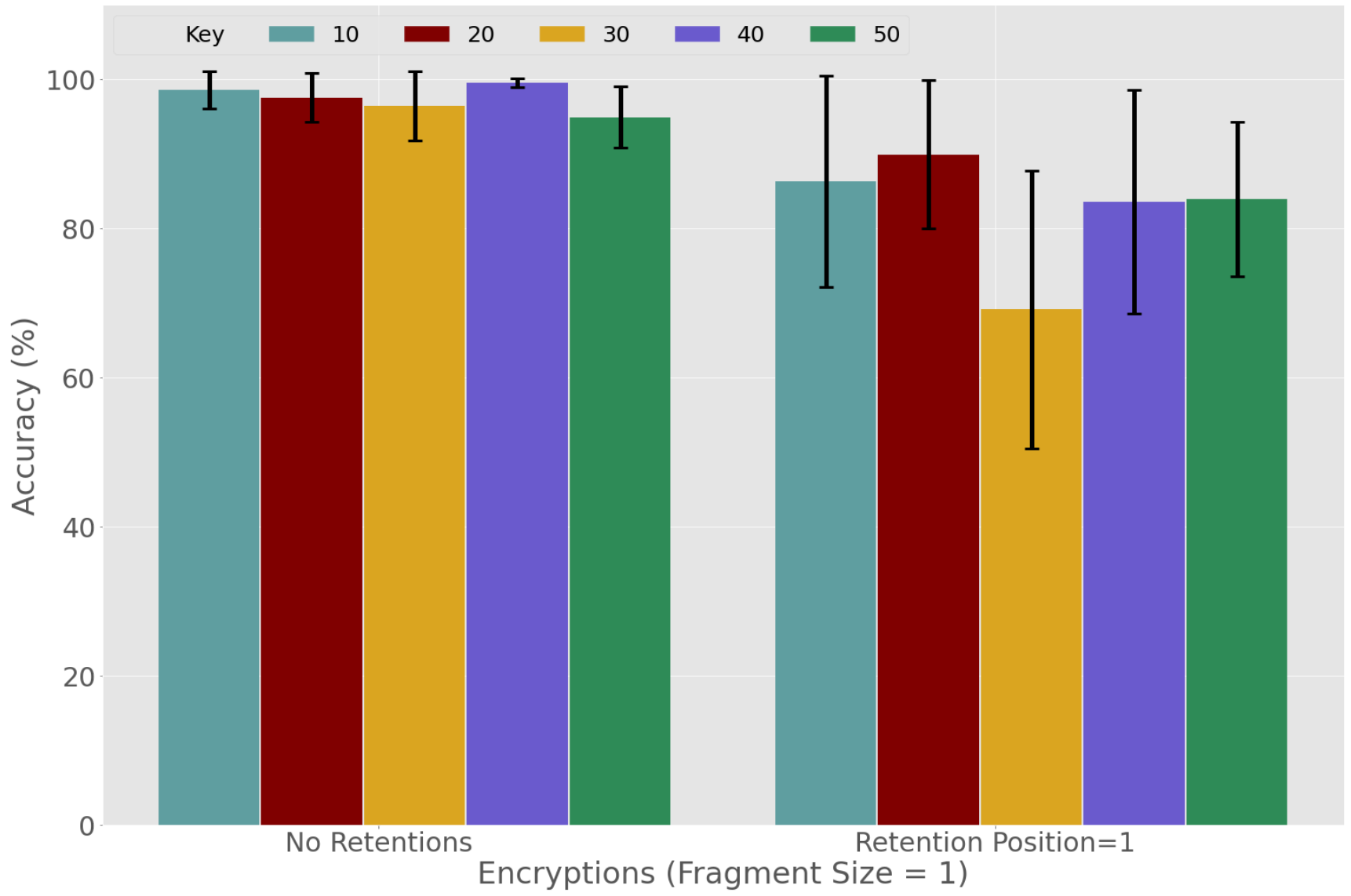

**Fig. 9:** *DNN-based detection of trigger samples amongst genuine E.Coli plasmids: the impact of nucleotide retention, no knowledge of the encryption key*

We will now further analyze the impact of encryption in detection. In **Fig. 9** we present the detection accuracies where the trojan payload address in the test data is encrypted with a different key. The model is trained with a particular key which is tested by all the data encrypted by the remaining keys. For example, the model trained by the data encrypted using key=10 will be tested by all the test data that are encrypted by other keys, i.e. keys = {20, 30, 40, 50}. Similarly, the model for key value 20 will be tested by all the test data encrypted by the keys = {10, 30, 40, 50}. In **Fig. 9** we plot the average accuracy against the different key values used for training the model. From this result, we conclude that a higher accuracy can be achieved for encrypted payloads without retention even if the key is unknown. However, the accuracy will deteriorate if we apply retention along with encryption. This is because the higher retention will result in the DNA sequence having a more natural pattern, which makes it more difficult to detect.

## Wet lab Experiments

In the previous sections of this article, we have described how we can disguise the address payload for a trojan attack to make the payload insert indistinguishable compared to a natural DNA sequence. Furthermore, applying encryption and steganography techniques will make it harder to detect the hybrid trojan attack. However, it is also important to address how practical it is to synthesize such a DNA sequence. In our wet-lab, we constructed the Trojan payload sequences both without and with encryption and steganography (**Fig. A.1** and **Fig. A.2**) via commercial gene synthesis with ease. These sequences were prepared and received already ligated into bacterial plasmid vector. These plasmids, pNOSTEG and pSTEG, were easily cloned into *E.coli* cells, propagated and purified in abundance (**Fig.A.3**). The Trojan payloads in both plasmids were both DNA sequenced completely and with 100% accuracy, with a sample chromatogram from pNOSTEG shown in **Fig. A.4**. We can assume that constructing natural DNA sequences will be easier and more achievable compared to synthesizing artificial DNA with unnatural sequences, due to possible runs and repeats of DNA bases that may cause problems in the synthesis reaction. As a result, there will be a need to construct a DNA that can allow multiple fragment inserts with the target information of the IP address and port number of the remote hacker's machine. With various techniques emerging for generating, producing or inserting multiple DNA sequences into carrier or expression systems, e.g., in-fusion cloning, gene assembly or multiple fragment cloning, hackers can bypass any gene synthesis issues by using a combination of these techniques to generate their final trojan attack sequence. As such, our work presents valuable detection against very feasible attack scenarios.

# Acknowledgements

This publication has emanated from research conducted with the financial support of Science Foundation Ireland (SFI) and the Department of Agriculture, Food and Marine on behalf of the Government of Ireland under Grant Number [16/RC/3835].

# Author Contribution

**Mr. Mohd Siblee Islam** is the primary author of the article. Mr. Islam was responsible for developing the software code used to perform computational experiment, executing the experiments, analysing and interpreting the results presented in this article, writing the manuscript.

**Dr. Stepan Ivanov** was responsible for overseeing and directing computational experiments presented in this article. Specifically, Dr. Ivanov contributed to the development of the proposed steganography technique, where he proposed the dynamic programming technique for finding an optimal location for the payload for malicious activity to be injected into the host DNA. Dr. Ivanov assisted Mr. Islam in writing the manuscript.

**Dr. Sasitharan Balasubramaniam** was the main scientific driver behind the experiments presented in the article. Due to his multidisciplinary background, Dr. Balasubramaniam identified the possibility for E.Coli bacteria to be used as carriers of malicious DNA on-purposed engineered as part of a Trojan attack. That was the starting point for the research presented in the article. Subsequently, Dr. Balasubramaniam directed and oversaw the experiments conducted in this research.

**Dr. Lee Coffey** planned and executed the wet lab experiments, including gene synthesis design, cloning and recombinant plasmid DNA purification.

**Dr. Srivatsan Kidambi** was responsible for providing expertise in methods for handling DNA based samples and background for DNA packaging/carrying.

**Ms. Jennifer Drohan** prepared the DNA samples for sequencing and carried out sequence analysis of the DNA fragments in order to verify sequence identity and fidelity.

**Dr. Witty Sri-saan** was the scientific driver behind the DNN analysis for the DNA strands with the injected code, as well as the development of the hacking scenarios.

**Dr. Hamdan Awan** was responsible for the analysis of the data in the results section and in particular the analysis on performance based on variations in parameters.

# Data Availability Statement

All data used in the manuscript are freely available in the public domain. The Programming code developed to conduct the experiments is freely available at the following URL: https://bitbucket.org/sibleeislam/bio-cyber-hacking.

## Competing Interests Statement

None of the authors or their respective organizations/research groups have any financial or otherwise interests that could affect or compromise findings of the research presented in this manuscript. The research presented in this article was carried out in strict accordance to the rules of research ethics and conduct.

## Artwork Statement

Artwork on Fig. 2 and 4 of the article was created by Mohd Siblee Islam using free Draw.io software and free icons available on the web. Artwork on Fig. 1 and 3 of the article was created by Dr. Ivanov and Mohd Siblee Islam using free Draw.io software and free icons available on the web.

## Supplementary Material: Appendix A: Generation of ‘steganography +/- ’ DNA

The actual content of the trojan payloads considered for the wetlab experiment is “a.wit:1753b.lab:8492”, which is also used as an example in our previous work [11]. This content is encoded into a DNA sequence with and without considering encryption and steganography. The algorithm used a fragment size of 4, key for the encryption of 60, and steganography key is 3. The encoded DNA sequences are shown in Fig. A.1 and Fig A.2.

***AGATATAAAGTACGACAGTGCTCTCGGCCCTT***
***AGATATACAGTACTCAATGGATACATCTCCTT***
***AGATATAGAGTAATCCATATCGAGAGTGCCTT***
***AGATATATAGTACGTACGACCGAGATGGCCTT***
***AGATATCAAGTAATGAATCAATGCATAGCCTT***

Fig. A.1: Without applying encryption and steganography - normal Trojan payload

***CTTATGAACGATTGTAATCAAGCAGGAATTCAAGTCTAGGTTTCAATTCTGTCTTCAAATG***
***TGCAGCTGGTCCTCGGACACAAAATGTTATGTTCAATATGCAGGCGTTACTCTAAAGATGA***
***TGCATG***
***CATCTCCCCTAATATTACGGAATTACTCCCAACATGTGAACGGACTGGCAGTCTTCGAGAT***
***TTTAACCGCATTGCCGCGGCTCGCGCCAAGGTCGCCTCGTATGCCTGGCATATAAAGGTCC***
***GAAAGG***
***ACTAGAATCGCCTGTACCGTCACTTCGCCATTAGCCTCCGTCGGGGTGAGAGAATAGTTAA***
***GCTACCTTCAATCAGCTGCCTCAAAACCGGTAATACACGGTGCCGGATTTGCTCAAATTAA***
***TTAGCT***
***CTTGTTATAGATGAGGCCGACCGGTAAAAATAATACTAGGTAACGTGCATTTCTGATGCTT***
***CATCGCAGAGCCATCTGAGAAAGCGGTGTCCGGATCGCTTCCCGGATCGGCCACCAACCTA***
***CCGGCG***
***CGTGGAATCCATGATCCCGTACTCCACTCTGAAGGTGTCGGTCTATCACGGCCGGGGAGCA***
***AACCGGATTACATGGATCTATTGCAAGTTAAACAGAGGCGGGTGGTCAACCACGACATATT***
***TGGAGG***

Fig. A.2: Trojan payload applying encryption and steganography

In these sequences, each line corresponds to a fragment of the trojan payload address (host names and port addresses only). We can insert any encoded line representing a fragment (without breaking) at any position inside an existing DNA sequence (also called our host DNA). However, note that we can not break a encoded line further as it represents a fragment. Furthermore the overlapping (if any) needs to be managed carefully. To summarize, the content of one file can be placed inside one plasmid, where any line can be put at any position (i.e., each line is a part of either the host name or port address of different machines that want to form a connection).

**Gene synthesis of DNA fragments with and without applying encryption and steganography**

Two separate samples were analyzed; both containing the Trojan payload enabling the attack, but one uses no encryption and steganography (normal Trojan payload) and another uses both encryption and steganography (Trojan payload applying steganography). The sample containing the Trojan payload with steganography, has a length of 640 basepairs (See **Fig. A.2**), while the sample containing the normal Trojan payload has 160 basepairs in length (See **Fig. A.1**). Both DNA sequences were gene synthesised by Eurofins Genomics Europe, Germany, provided in pEX-A128A plasmid vector in lyophilised format and the resulting plasmids named pSTEG and pNOSTEG respectively.

**Preparation and transformation of competent cells**

*E.Coli* NovaBlue cells (Novagen, *endA1 hsdR17*($r_{K12}^{-}m_{K12}^{+}$) *supE44 thi-1 recA1 gyrA96 relA1 lac* F'[proA$^{+}$B$^{+}$ *lacI*$^{q}$ *Z∆M15*::Tn*10* (Tc$^{R}$)]) were inoculated into LB broth and grown overnight at 37°C in a shaking incubator at 250 RPM with adequate aeration. Competent cells were then prepared using the *Mix&Go! E. Coli* Transformation Kit (Zymo Research) as per manufacturer's instructions.

Transformation of competent *E. Coli* NovaBlue cells with pEX-A128 'Trojan payload applying steganography' (pSTEG) and 'Normal Trojan Payload' (pNOSTEG) plasmid DNA was carried out by adding 1 µL of the relevant resuspended plasmid DNA into 50 µL of competent *E. coli* NovaBlue cells, as per *Mix&Go!* kit protocol, and aliquots were spread on pre-warmed LB/Amp (Ampicillin 100 µg/ml) agar plates. A negative control plate was prepared by adding 1 µL of sterile water in place of DNA. Plates were incubated at 37°C overnight. Successfully transformed cells were selected via ampicillin resistance as a selection marker.

Successfully transformed isolated colonies were then inoculated into LB/Amp broth and cultures were incubated until an $OD_{600nm}$ = 2 was reached. $OD_{600nm}$ measurements were taken using the NanoDrop™ 1000 (Thermo Scientific™). Cultures were then concentrated to an $OD_{600nm}$ = 10. Once cultures were at the appropriate $OD_{600nm}$, plasmid DNA was purified using the Monarch® Plasmid Miniprep Kit (NEB) as per manufacturer's instructions. Plasmid samples were eluted in sterile water and the DNA concentration and quality was assessed using the NanoDrop™ 1000. The presence of the plasmid for each sample was verified using agarose gel electrophoresis (0.8% agarose made with 1xTAE buffer) (**Fig. A.3**).

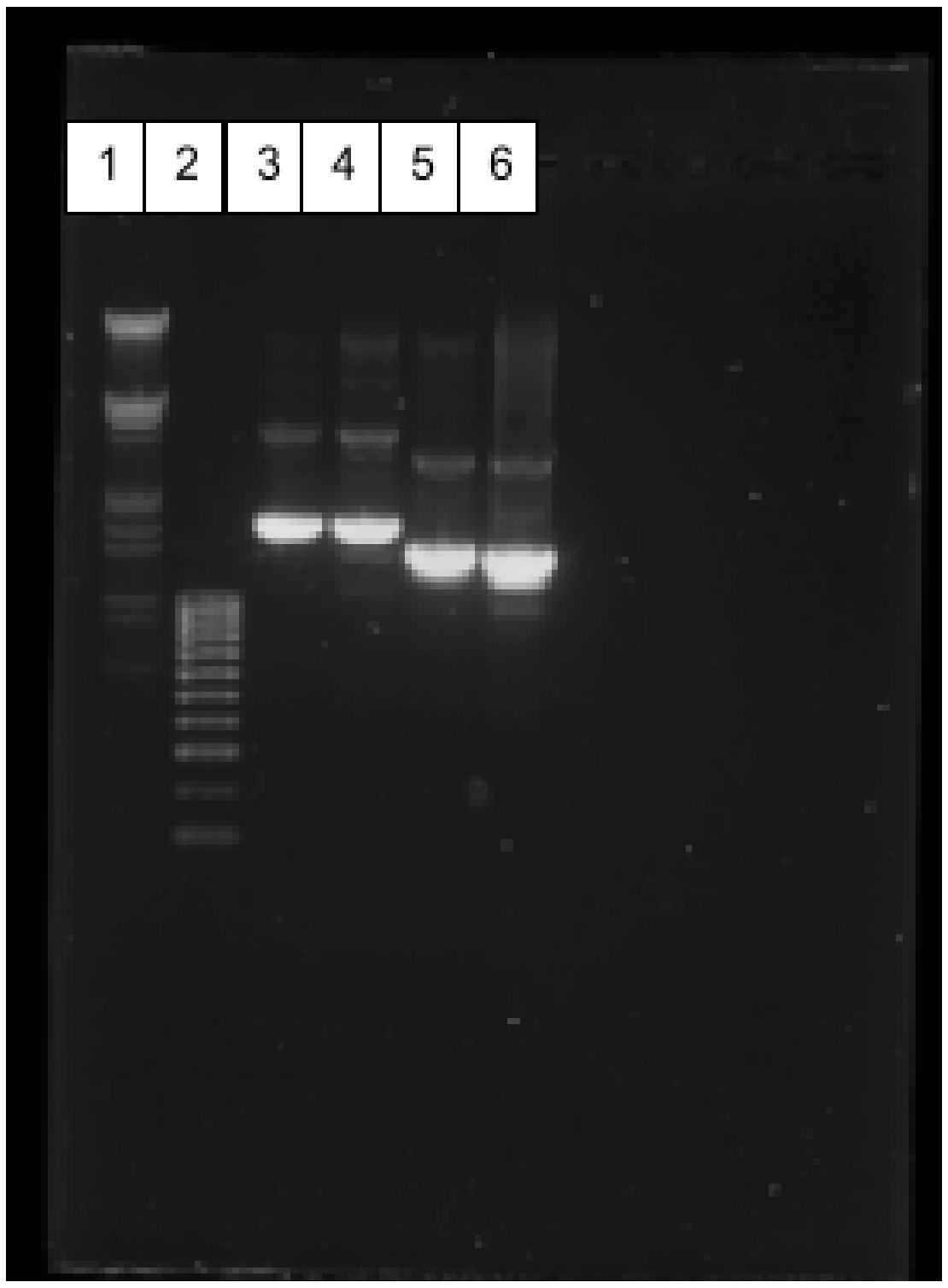

**Fig. A.3:** Confirmation of plasmid purification using agarose gel electrophoresis. 1: Lambda DNA/EcoR1 plus HindIII ladder. 2: Promega 100bp ladder. 3+4: pSTEG plasmid - Trojan payload applying steganography 5+6: pNOSTEG plasmid - normal Trojan payload.

## DNA Sequencing

Samples were sequenced by Eurofins Genomics Europe Sequencing GmbH, Germany. Oligonucleotides used for sequencing were supplied by Eurofins; pEX-For (5'-GGAGCAGACAAGCCCGTCAGG-3') and pEX-Rev (5'-CAGGCTTTACACTTTATGCTTCCGGC-3').

## Analysis of sequencing data

Analyses of sequencing data were carried out using a combination of Chromas (v 2.6.6) and MEGA-X (v 10.2.6). Sequencing chromatogram quality was first assessed using Chromas. Sequence alignments were performed using the CLUSTALW algorithm in MEGA-X. Following successful alignment of DNA sample sequence with reference sequence, the sequences were trimmed in Chromas to highlight the 'Trojan payload applying steganography' DNA and 'Normal Trojan payload' DNA only for analysis. (Sample sequencing results are shown in Fig A.4).

## Sequencing Results

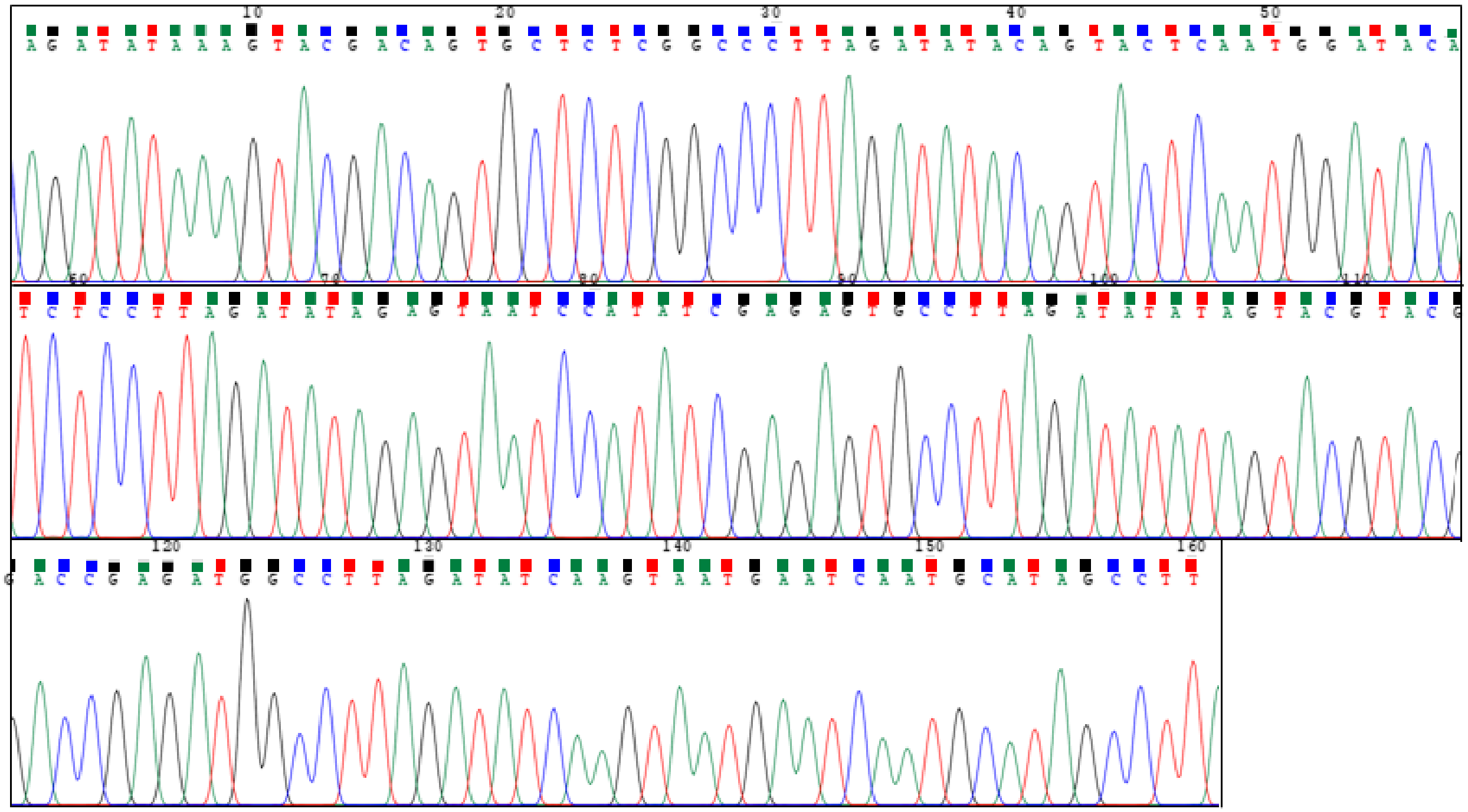

**Fig. A.4:** Sample sequencing chromatogram from pNOSTEG with 60bp DNA sequence region for Trojan payload address without encryption and steganography applied visible.